\documentstyle[prb,twocolumn,floats,epsf,aps]{revtex}
\begin{document}
\draft

\twocolumn[\hsize\textwidth\columnwidth\hsize\csname @twocolumnfalse\endcsname

\title{Ground state parameters, finite-size scaling, and low-temperature
properties of the two-dimensional $S=1/2$ XY model}

\author{Anders W. Sandvik}
\address{Department of Physics, University of Illinois at Urbana-Champaign,
1110 West Green Street, Urbana, Illinois 61801 \\
and Center for Nonlinear Studies, Los Alamos National Laboratory,
Los Alamos, New Mexico 87545}

\author{Chris J. Hamer}
\address{School of Physics, University of New South Wales, Sydney
2052, Australia}

\date{\today}

\maketitle

\begin{abstract}
We present high-precision quantum Monte Carlo results for the $S=1/2$
XY model on a two-dimensional square lattice, in the ground state as well 
as at finite temperature. The energy, the spin stiffness, the magnetization,
and the susceptibility are calculated and extrapolated to the thermodynamic
limit. For the ground state, we test a variety of finite-size scaling 
predictions of effective Lagrangian theory and find good agreement and
consistency between the finite-size corrections for different quantities. 
The low-temperature behavior of the susceptibility and the internal 
energy is also in good agreement with theoretical predictions.
\end{abstract}


\vskip2mm]

\section{Introduction}

Studies of effective continuum field  theories have resulted in detailed 
predictions for the low-energy physics of quantum spin systems in two 
dimensions.\cite{csh,neuberger,fisher,hasenfratz} In order to make use of 
these predictions for a given model Hamiltonian, the ground state 
parameters appearing in the Lagrangian formulation have to be determined. 
Spin-wave theory \cite{spinwave1} can in some cases give good estimates,
\cite{spinwave2} but in general some numerical method has to be 
employed in order to obtain accurate results. Since the theories can 
also predict the finite-size scaling behavior of various physical 
quantities, numerical results for a series of lattice sizes can be used 
to extract the ground state parameters. Such calculations are also 
important for testing theoretical predictions. 

With todays computers, Lanczos and related exact diagonalization methods
can be used for square lattices with up to $6 \times 6$ spins.
\cite{schulz,einarsson,hamer} 
This relatively small maximum size, and the small number of different lattices 
available, can make the finite-size scaling procedures problematic if
sub-leading corrections are significant. It is therefore important to 
consider also alternative methods that can reach larger lattice sizes.

Here we discuss quantum Monte Carlo (QMC) results for the $S=1/2$ XY model,
defined by the Hamiltonian
\begin{equation}
H = -J\sum\limits_{\langle i,j\rangle} [S^x_iS^x_j  + S^y_iS^y_j],
\label{model}
\end{equation}
where $S^x_i$ and $S^y_i$ are the $x$- and $y$-components of a spin-1/2
operator at site $i$, and $\langle i,j\rangle$ denotes a pair of
nearest-neighbor sites on a square lattice. Numerical studies of this
model have a long history.\cite{pearson,oitmaa,loh,okabe,zhang,harada,hamer} 
Exact finite-lattice calculations gave the first indications that the
O(2) symmetry is spontaneously broken at $T=0$ in the thermodynamic limit.
\cite{oitmaa} This was later proved rigorously.\cite{orderproof} QMC 
simulations have resulted in a quite precise value of the ordered moment 
\cite{zhang} and have also shown \cite{loh} that there is a 
Kosterliz-Thouless (KT) transition at a temperature 
$T_{\rm KT} \approx 0.343J$.
\cite{harada} For $T \le T_{\rm KT}$ the system acquires quasi-long-range 
order, i.e., power law decay of the spin-spin correlation function and 
a non-zero spin stiffness constant. In accordance with the Mermin-Wagner 
theorem,\cite{mermin} true long-range order develops only at $T=0$.

Our motivation for carrying out calculations for the XY model to even 
higher precision is to test predictions of effective Lagrangian theories.
In particular, the chiral perturbation calculations by Hasenfratz and 
Niedermayer have resulted in detailed predictions for the finite-size and 
finite-temperature corrections of various quantities,\cite{hasenfratz} 
in some cases beyond leading order. The finite-size and finite-temperature
scaling behavior of the O(3) symmetric Heisenberg model has been the 
topic of numerous studies,\cite{schulz,runge,wiese,troyer,sandvik}
and the agreement with the predictions has been confirmed to high 
precision. The predictions for the O(2) symmetric XY model have not yet 
been tested exhaustively, however. A recent finite-size scaling 
study of exact energies for systems with up to $6 \times 6$ spins has quite
convincingly demonstrated agreement with the leading finite-size 
behavior.\cite{hamer} Here we consider also several other physical 
observables. This allows us to carry out a number of independent tests 
of the consistency of the scaling predictions. With access to larger 
system sizes, we can also improve considerably on the accuracy of the 
extrapolated ground state parameters and finite-size corrections. Our 
data are sufficiently accurate for addressing also sub-leading corrections.
In addition to calculations in the ground state, we have also studied 
systems at finite temperature, on lattices sufficiently large to enable 
extraction of the leading finite-temperature corrections in the 
thermodynamic limit. 

We have used a numerically exact finite-temperature QMC method based on 
``stochastic series expansion'',\cite{sse} i.e., importance sampling 
of diagonal matrix elements of the power series expansion of 
${\rm exp}(-\beta H)$, where $\beta=1/T$ is the inverse temperature. 
The ground state can be obtained by choosing a sufficiently large 
$\beta$ value. This method has previously been used to study the 
finite-size scaling behavior and the ground state parameters of the 
square-lattice Heisenberg model.\cite{sandvik} Recently, a new way of 
sampling the terms of the expansion was proposed (the ``operator-loop'' 
algorithm),\cite{loop} which enables calculations to even higher 
precision. We have used this improved method to study the ground state 
of the XY model on lattices with $N=L\times L$ spins with $L$ up to 
$16$. We have carried out calculations at finite temperature for 
systems with $L$ up to $64$. 

In Sec.~II we discuss the physical quantities that we have calculated.
The numerical ground state data and the results for finite-size 
corrections and extrapolations to infinite size are presented in 
Sec.~III. In Sec.~IV we discuss calculations at $T > 0$. Sec.~V 
concludes with a brief summary.

\begin{table}
\begin{tabular}{lddd}
 $L$ &  ~~~~$-E_0$  & ~~~~$\rho$  & ~~~~~~~~$M_x^2$   \\ \hline
  4  & 0.562485(4)   & 0.2769(1)  & 0.13282(2)  \\
  6  & 0.552696(4)   & 0.2718(1)  & 0.11885(4)  \\
  8  & 0.550436(4)   & 0.2705(2)  & 0.1126(2)   \\
 10  & 0.549643(4)   & 0.2700(3)  & 0.1087(2)   \\
 12  & 0.549296(4)   & 0.2698(4)  & 0.1065(3)   \\
 14  & 0.549118(4)   & 0.2695(3)  &     \\
 16  & 0.549020(4)   & 0.2699(4)  &     \\
\end{tabular}
\vskip2mm
\caption{QMC results for the ground state energy, the spin stiffness,
and the squared magnetization per spin. The numbers within parentheses
indicate the statistical errors of the least significant digit of the 
results.}
\label{tab1}
\end{table}

\begin{figure}
\centering
\epsfxsize=8cm
\leavevmode
\epsffile{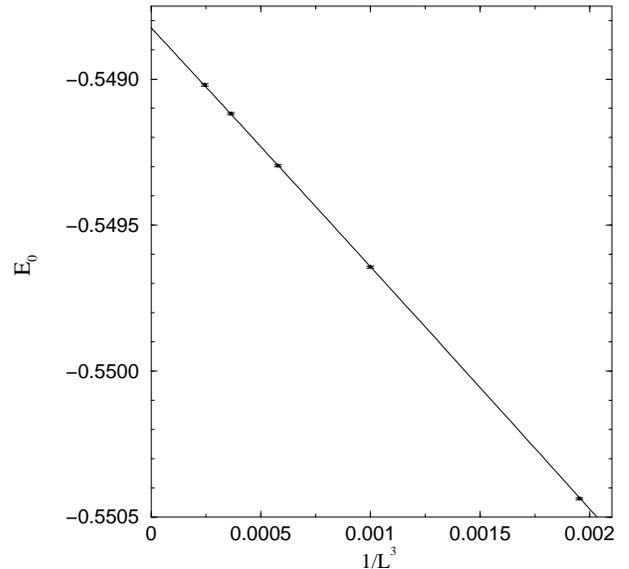}
\vskip1mm
\caption{Ground state energy vs the inverse cubed of the system 
size for $L=8,10,12,14$, and $16$ (points with error bars). The
curve is a fit to Eq.~(\protect{\ref{e0scale}}), including also $L=4$
and $L=6$.}
\label{fige0}
\end{figure}

\section{Calculated quantities}

The QMC algorithm \cite{loop} can be implemented in two different ways;
with the spin quantization either along the $z$ or $x$ axis. The internal
energy, $E=\langle H\rangle /N$,
can be easily calculated in both cases. In the $z$-representation, 
the spin stiffness constant $\rho$ can also be evaluated. It is defined 
as the second derivative of the internal energy with respect to a twist 
$\phi$ under which, for all nearest-neighbor spin pairs in either the 
$x$ or $y$ lattice direction, the spin-spin interaction is modified 
according to
\begin{eqnarray}
S^x_iS^x_j && + S^y_iS^y_j \to
(S^x_iS^x_j + S^y_iS^y_j)\cos{(\phi)} \nonumber \\
&& - S^x_iS^y_j\sin{(\phi)} + S^y_iS^x_j\sin{(\phi)}.
\end{eqnarray}
In analogy with the superfluid density of a boson system,\cite{pollock} 
one can show that\cite{sandvik}
\begin{equation}
\rho = {\partial ^2 E(\phi) \over \partial^2 \phi } = 
{\langle W_x^2 \rangle + \langle W_y^2 \rangle \over 2N\beta},
\end{equation}
where $E(\phi)$ is the internal energy per spin and $W_x$ and $W_y$ are 
the ``winding numbers'', i.e., the net spin currents across the
periodic boundaries in the $x$ and $y$ directions that characterize 
configurations in simulations carried out in the $z$-representation. 
We also calculate the spin susceptibility, given by
\begin{equation}
\chi = {\beta\over N^2} 
\left\langle \left ( \sum\limits_{i=1}^N S^z_i\right )^2 \right\rangle .
\label{suscept}
\end{equation} 
In the $z$-representation,
it is not convenient to calculate spin-spin correlations involving
the $x$ or $y$ spin components. We therefore also use an
algorithm implemented in the $x$ representation, and there calculate 
the squared magnetization $M_x^2$;
\begin{equation}
M_x^2 = {1\over N^2} \left\langle \left ( \sum\limits_{i=1}^N S^x_i 
\right ) ^2 \right\rangle .
\end{equation}
Since the O(2) symmetry is not broken on a finite lattice, the magnetization
$m$ per spin in the thermodynamic limit is given by the infinite-size 
extrapolated value of
\begin{equation}
m=\sqrt{2M_x^2} 
\label{mM}.
\end{equation}

We use periodic boundary condition and include each spin pair only once 
in Eq.~(\ref{model}). For technical details on the simulation
algorithm we refer to previously published work.\cite{loop,sandvik}

\section{Ground state results}

In order to obtain the ground state, we have carried out simulations
at inverse temperatures as high as $\beta=1024$ for lattices with $L=4-16$.
In the $z$-representation simulations, we have calculated the ground 
state energy to within relative statistical errors lower than $10^{-5}$. 
The relative accuracy of the stiffness is on the order of $10^{-3}$. 
Simulations in the $x$-representation result in statistical errors for 
the energy roughly twice as large as in the $z$-representation (for 
simulations of comparable length). The squared magnetization is the 
least accurate quantity, with statistical errors of roughly $0.3\%$ 
for the largest system size considered in this case ($L=12$). 
In Table \ref{tab1} we list the results used in the finite-size 
analysis presented below. The results for the
energy for $L=4$ and $6$ are in perfect agreement with previous 
\cite{oitmaa,hamer} exact diagonalizations. The magnetization for
$L=4$ also agrees with the exact diagonalization result. However,
for $L=6$ there is an $\approx 5\sigma$ deviation from the result
presented in Ref.~\onlinecite{hamer}. We do not know the reason
for this disagreement. The QMC method has previously been shown to 
give perfect agreement with isotropic Heisenberg results for both
$L=4$ and $L=6$,\cite{sandvik} and therefore a failure for (only) the
XY-model magnetization for $L=6$ would be surprising.

\begin{figure}
\centering
\epsfxsize=8cm
\leavevmode
\epsffile{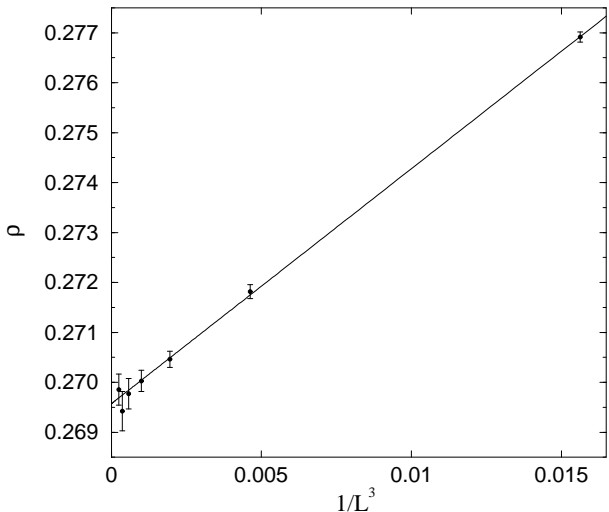}
\vskip1mm
\caption{Spin stiffness vs the inverse cubed of the system size. 
The line is a fit to the cubic form (\protect{\ref{rho0scale}}).}
\label{figrho0}
\end{figure}

For a model with O(2) symmetry, chiral perturbation theory predicts
the size dependence of the energy as \cite{hasenfratz}
\begin{equation}
E_0(L) = E_0 + {e_3 \over L^3}  + {e_5 \over L^5} + \ldots,
\label{e0scale}
\end{equation}
with no $O(1/L^4)$ term. The leading-order correction is given
by 
\begin{equation}
e_3 = -\gamma c/2 ,
\label{e3}
\end{equation}
where the constant $\gamma=1.437745$ and $c$ is the spin-wave velocity. 
In order to obtain a good fit to our $L=4-16$ data, the $O(1/L^5)$ term
has to be included. The fit then has a chi-squared value per degree of 
freedom of $0.7$. Using instead an $O(1/L^4)$ term gives chi-squared 
$\approx 6$. We can therefore conclude that our results 
support the prediction \cite{hasenfratz}
that $e_4=0$. In Figure \ref{fige0} we show the
data for $L \ge 8$ along with the best fit. The extrapolation 
$L \to \infty$ gives $E_0 = -0.548824(2)$. The finite-size correction
constants are $e_3 = -0.807(2)$ and $e_5 = -1.07(3)$. Using
Eq.~(\ref{e3}) we obtain the spin-wave velocity $c=1.123(2)$. These
results are in agreement with the previous extrapolations using
exact diagonalization data for $L=2,3,4,5,6$,\cite{hamer} but the statistical
errors are considerably smaller. The energy is also in perfect agreement
with a previous Green's function Monte Carlo calculation, which gave
$E_0 = 0.54883(1)$.\cite{zhang}

We are not aware of any predictions for the size dependence of
the spin stiffness of the XY model. For the Heisenberg model, the 
leading correction is $O(1/L)$.\cite{einarsson} In contrast, our XY 
data can be very well fit with only an $O(1/L^3)$ term;
\begin{equation}
\rho (L) = \rho + {r_3 \over L^3} + \ldots,
\label{rho0scale}
\end{equation}
with the infinite-size value $\rho=0.2696(2)$ and the cubic correction
$r_3 = 0.47(1)$. This fit is shown in Figure \ref{figrho0}. 

\begin{figure}
\centering
\epsfxsize=8cm
\leavevmode
\epsffile{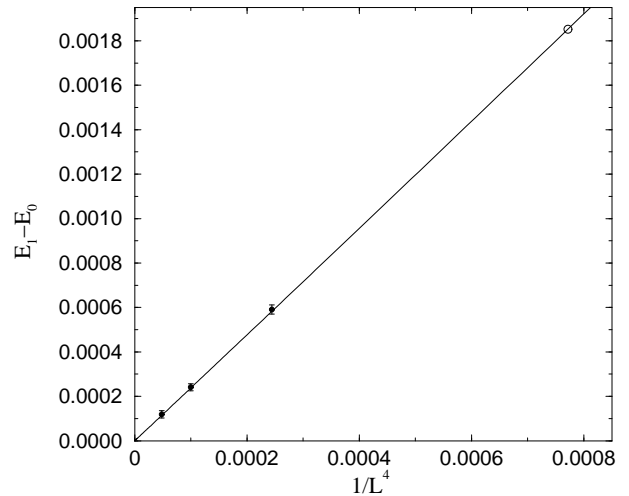}
\vskip1mm
\caption{The difference between the energy in the $j=1$ and $j=0$
magnetization sectors graphed vs $1/L^4$. The open circle is the exact 
result for $L=6$. The points with error bars are the QMC results for 
$L=8,10$ and $12$. The curve is a fit to Eq.~(\protect{\ref{eje0}}).}
\label{figej}
\end{figure}

The stiffness result
is in perfect agreement with the value obtained using exact diagonalization
and  finite-size scaling of the ground state energy
in higher magnetization sectors.\cite{hamer} The energy per spin in the sector
with magnetization $j=\sum_iS^z_i$ should scale as \cite{hasenfratz}
\begin{equation}
E_j - E_0 = {j^2c^2 \over 2\rho L^4} + O(j^4,1/L^6).
\label{eje0}
\end{equation}
We have also carried out some QMC calculations in the $j=1$ sector. Figure
\ref{figej} shows results for $L=8,10$, and $12$, along with the exact 
result for $L=6$. Using also the exact $L=4$ result, a fit to 
Eq.~(\ref{eje0}), including an $O(1/L^6)$ term, gives $c^2/\rho = 4.70(4)$. 
This value agrees with the above separate estimates of 
$c$ and $\rho$. The consistency between the results, obtained in two 
different ways, clearly gives very strong support to the 
Lagrangian theory.

The susceptibility can be obtained from $c$ and $\rho$ using
the standard hydrodynamic relation
\begin{equation}
\chi = \rho / c^2 ,
\end{equation}
which with our values of $c$ and $\rho$ gives $\chi = 0.2138(8)$.

\begin{figure}
\centering
\epsfxsize=8cm
\leavevmode
\epsffile{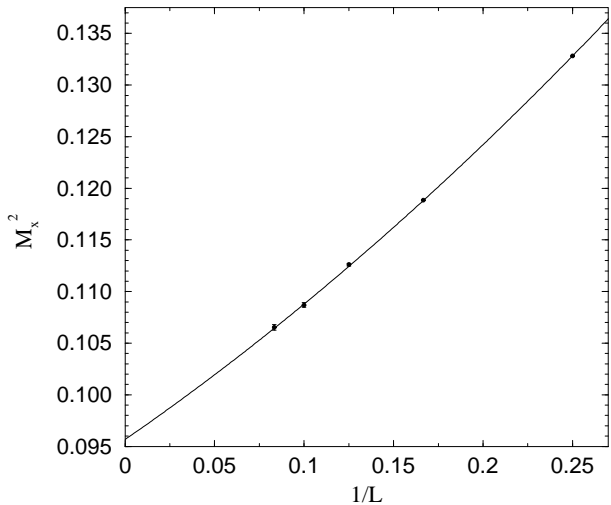}
\vskip1mm
\caption{Magnetization vs inverse system size and a fit to the
scaling form (\protect{\ref{scalem}}).}
\label{figm}
\end{figure}

For the magnetization, we assume a finite-size scaling
\begin{equation}
M^2_x(L) = M^2_x + {a_1 \over L} + {a_2 \over L^2} + \ldots .
\label{scalem}
\end{equation}
A fit to this form, including the quadratic term, is shown in Figure 
\ref{figm} and gives $M_x^2 = 0.0956(6)$ and $a_1 = 0.1212(5)$. According to
Eq.~(\ref{mM}), the magnetization is thus $m=0.437(2)$, in good agreement
with a previous Green's function Monte Carlo result [$m=0.441(5)$].\cite{zhang}
For the Heisenberg model, the linear correction factor is related to ground
state parameters according to $a_1 = \alpha M^2_x /(c\chi)$, where 
$\alpha = 0.62075$.\cite{neuberger} The $a_1$ obtained here is instead 
consistent to within a few percent with $a_1 = \alpha M^2_x /(2c\chi)$.
As in the scaling of the energy,\cite{hasenfratz} the leading size 
correction is hence proportional to the number of gapless modes 
in the symmetry-broken system.

\section{Finite-temperature results}

We now discuss calculations aimed at extracting the leading finite-temperature
corrections to the ground state. Previously, extensive simulations were
carried out in order to study the KT transition, and a critical temperature
$T_{\rm KT} \approx 0.343J$ was found.\cite{harada} Here we focus on the 
susceptibility and the internal energy density at lower temperatures.

\begin{figure}
\centering
\epsfxsize=8cm
\leavevmode
\epsffile{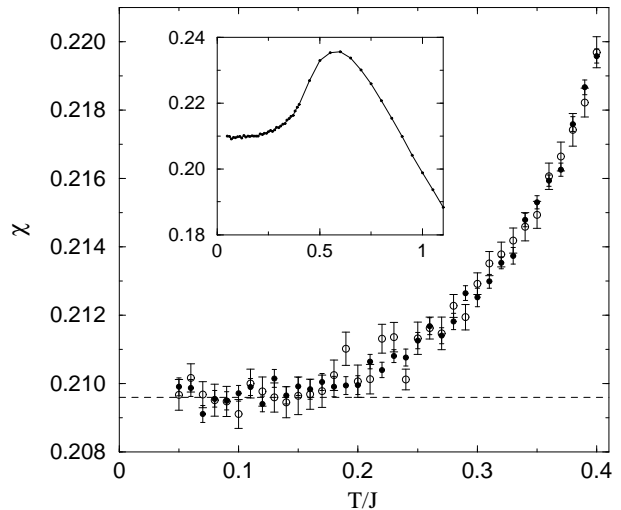}
\vskip1mm
\caption{Spin susceptibility vs temperature for lattices with $L=32$
(solid circles) and $L=64$ (open circles). The dashed line is the
estimated constant low-temperature value. The inset shows results
for $L=32$ over a wider temperature range.}
\label{figxt}
\end{figure}

Figure \ref{figxt} shows the susceptibility, defined in Eq.~(\ref{suscept}), 
calculated for lattices with $L=32$ and $64$ for temperatures $T/J \ge 0.05$. 
Within statistical errors there are no differences between the data for
the two system sizes, and hence the results represent the thermodynamic 
limit. Chiral perturbation theory predicts a temperature-independent 
susceptibility for the XY model.\cite{hasenfratz} Within error bars,
our data is temperature independent for $T/J \alt 0.15$, with the 
constant value estimated at $\chi = 0.2096(2)$. This result is 
lower (by about five standard deviations, or $2$\%) than the value 
extracted in Sec~III from the ratio $\chi=\rho/c^2$. The spin-wave 
velocity was obtained from the 
leading finite-size correction to the ground state energy. It is 
possible that this correction is affected by the presence of sub-leading 
corrections that are not completely taken into account by only including 
up to $O(1/L^5)$ terms in Eq.~(\ref{e0scale}). Using the data shown in 
Fig.~\ref{figxt} is a more direct way of extracting the $T=0$ 
susceptibility, and we therefore consider it more reliable. An
improved estimate of $c$ is then $c=\rho/\chi =1.134(2)$.

\begin{figure}
\centering
\epsfxsize=8cm
\leavevmode
\epsffile{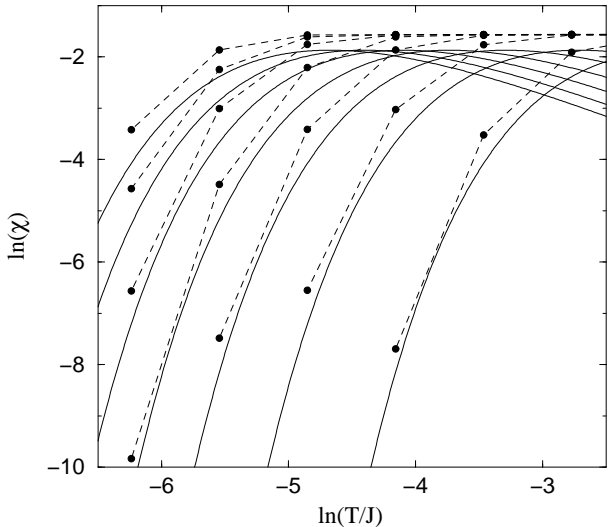}
\vskip1mm
\caption{The logarithm of the spin susceptibility for different 
system sizes vs the logarithm of the temperature (solid circles
connected by dashed lines, $L=4,6,\ldots,16$ from right to left).
The solid curves are obtained from Eq.~(\protect{\ref{xltprediction}})
with $c^2/\rho = 4.77$.}
\label{figxlt}
\end{figure}

\begin{figure}
\centering
\epsfxsize=8cm
\leavevmode
\epsffile{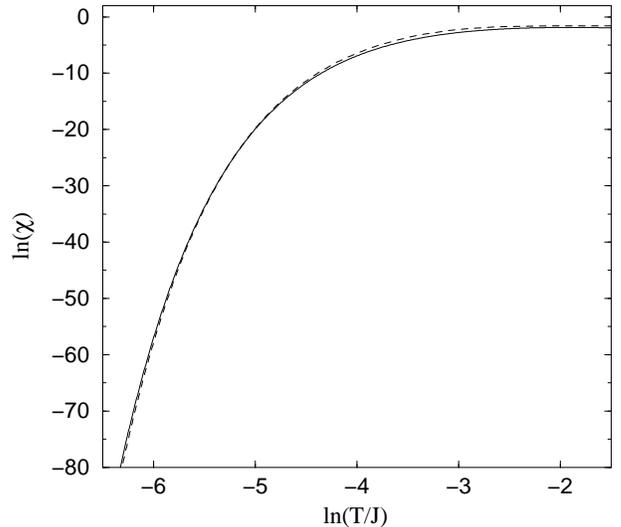}
\vskip1mm
\caption{The logarithm of the exact spin susceptibility for a $4\times 4$
lattice vs the logarithm of the temperature (dashed curve) compared
with the predicted form (\protect{\ref{xltprediction}}) with  
$c^2/\rho = 4.77$ (solid curve).}
\label{figxt4}
\end{figure}

For a finite lattice, the gap between the $j=0$ ground state and
the finite-magnetization states implies an exponential decay of $\chi$
to zero as $T \to 0$. Chiral perturbation theory predicts that the 
form of this decay for $TL^2\rho/c^2 \ll 1$ is given by \cite{hasenfratz}
\begin{equation}
\chi = {1 \over TL^2} {\rm exp}[-c^2/(2\rho TL^2)].
\label{xltprediction}
\end{equation}
In Figure~\ref{figxlt} we show  QMC data for $L=4-16$ along with this
prediction, where we have used $c^2/\rho=4.77$, corresponding to our best
estimates of $c$ and $\rho$. The agreement is not perfect, but satisfactory
considering that there are no adjustable parameters and that there
should also be corrections to Eq.~(\ref{xltprediction}). For $L=4$,
we have also calculated $\chi$ using exact diagonalization down to much
lower temperatures. As shown in Figure \ref{figxt4}, there is a very good
agreement with the predicted form over a sizable low-temperature range. 
Note, however, that the asymptotic  $T \to 0$ decay is always purely 
exponential, without the $1/T$ factor in Eq.~(\ref{xltprediction}),
as only the lowest $j=\pm 1$ states contribute.

\begin{figure}
\centering
\epsfxsize=8cm
\leavevmode
\epsffile{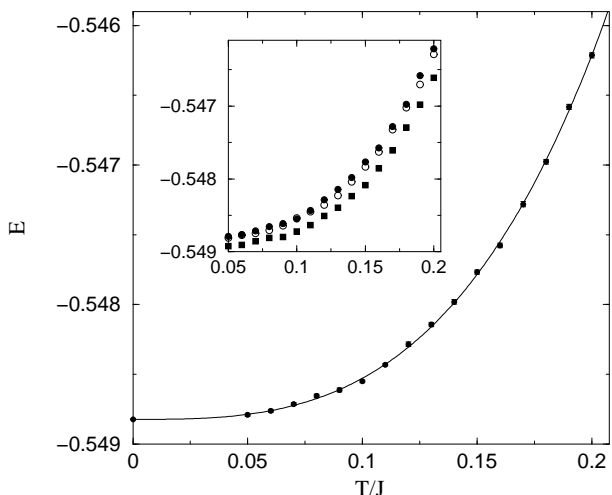}
\vskip1mm
\caption{The internal energy calculated for $L=64$ lattices, and a fit 
to the form (\protect{\ref{etscale}}). The inset shows data for $L=16$
(squares), $32$ (open circles), and $64$ (solid circles).}
\label{figet}
\end{figure}

Chiral perturbation theory predicts that the low-temperature form of the
internal energy is given by\cite{hasenfratz}
\begin{equation}
E(T) = E(0) + {\zeta (3) \over \pi c^2}T^3  + O(T^5),
\label{etscale}
\end{equation}
where $\zeta (3) = 1.20206$. In Figure \ref{figet} we show results for
$T/J \ge 0.05$ calculated for $L=64$ lattices. We also show a comparison 
between $L=16,32$, and $64$, which suggests that $L=64$ gives the 
thermodynamic limit within error bars. A fit to the $L=64$ data with 
$O(T^3)$ and $O(T^5)$ terms, and including also the $T=0$ energy extracted
in Sec.~III, gives the cubic correction $\zeta (3)/ \pi c^2=0.284(5)$. This 
corresponds to $c=1.16(1)$, which is higher than the value $1.134(2)$ 
extracted above using the estimates of $\rho$ and $\chi$ by about 2.5 
error bars. Again, this small deviation may indicate some influence of 
higher-order corrections  in the energy scaling.

\section{Summary}

In summary, we have presented extensive QMC calculations for the 
two-dimensional $S=1/2$ XY model. We have carried out finite-size and 
finite-temperature scaling of several physical quantities. The results 
are consistent with the predictions of effective Lagrangian theory 
\cite{hasenfratz} to within $1-2$\%. 

The best estimates of the ground state parameters resulting from
our calculations are
\begin{eqnarray}
E_0   & = & 0.548824(2), \nonumber \\
\rho  & = & 0.2696(2), \nonumber \\
m     & = & 0.437(2), \nonumber \\
\chi  & = & 0.2096(2), \nonumber \\
c     & = & 1.134(2). \nonumber 
\end{eqnarray}
The ground state energy, $E_0$, the spin stiffness, $\rho$, and the
square of the magnetization, $m^2$, were all calculated directly
in the ground state for systems of linear size $L=4-16$ and extrapolated
to infinite size. The susceptibility $\chi$ perpendicular to the XY
spin-plane was calculated at finite temperature and extrapolated to $T=0$.
Only the spin-wave velocity $c$ was obtained by a more indirect procedure,
using the relation $c^2 = \rho/\chi$. The results are in good agreement
with previous exact diagonalization \cite{oitmaa,hamer} and QMC
work,\cite{zhang} but the precision is considerably improved over 
other estimates. Our results for both the energy and the magnetization
are in remarkably good agreement with a series expansion calculation.
\cite{hamer2} The magnetization is also in excellent agreement with
second order spin-wave theory.\cite{hamer2} The energy obtained in 
spin-wave theory \cite{hamer2} is only $0.2$\% higher than the
numerical result obtained here. It would be interesting to carry 
out high-order spin-wave calculations also for the other quantities
discussed here.

\section{Acknowledgments}

We would like to thank Jaan Oitmaa for discussions and comments on the
manuscript. A. W. S. would like to thank the School of Physics at the 
University of New South Wales for hospitality and financial support 
during a visit. Support from the NSF under Grant No.~DMR-9712765 
is also acknowledged. 

\end{document}